\documentclass[12pt]{article}
\usepackage{cite}
\usepackage{mathptmx}
\usepackage{setspace}

\usepackage{mathrsfs}

\usepackage{changepage}
\usepackage{dcolumn}
\usepackage[table]{xcolor}
\usepackage{floatrow}
\usepackage{floatrow}   
\usepackage{subcaption}
\usepackage{graphicx,xcolor} 
\usepackage{url}
\usepackage{authblk}
\usepackage{hyperref}

\usepackage[margin=1in]{geometry}
\usepackage{amsmath}
\usepackage{amssymb}
\usepackage{pgf}
\usepackage{comment}

\definecolor{mynavy}{HTML}{000080}
\definecolor{darkred}{HTML}{8B0000}
\definecolor{mygreen}{HTML}{006400}
\definecolor{mygold}{HTML}{B8860B}

\newcolumntype{d}[1]{D..{#1}}

\title{Income inequality and mobility in geometric Brownian motion with stochastic resetting: theoretical results and empirical evidence of non-ergodicity}

\author{ Viktor Stojkoski$^{1,2}$\footnote{These authors contributed equally to this work.}, Petar Jolakoski$^{3*}$, Arnab Pal$^{4}$, Trifce Sandev$^{2,5,6}$, \\ Ljupco Kocarev$^{2,7}$, Ralf Metzler$^{5\footnote{Corresponding author: \href{mailto:rmetzler@uni-potsdam.de}{rmetzler@uni-potsdam.de} }}$}

\affil{%
\footnotesize
$^{1}$Faculty of Economics, Ss.~Cyril and Methodius University, 1000 Skopje, Macedonia \\
\footnotesize
$^{2}$Research Center for Computer Science and Information Technologies, Macedonian Academy of Sciences and Arts \\\footnotesize
$^{3}$Association for Research and Analysis - ZMAI, 1000 Skopje, Macedonia \\
\footnotesize
$^{4}$Department of Physics, Indian Institute of Technology, Kanpur, Kanpur 208016, India \\
\footnotesize
$^{5}$Institute of Physics \& Astronomy, University of Potsdam \\\footnotesize
$^{6}$Institute of Physics, Faculty of Natural Sciences and Mathematics,
Ss Cyril and Methodius University\\\footnotesize
$^{7}$Faculty of Computer Science and Engineering, Ss. Cyril and Methodius University}

\begin{document}
\maketitle
\begin{abstract}
We explore the role of non-ergodicity in the relationship between income inequality, the extent of concentration in the income distribution, and mobility, the feasibility of an individual to change their position in the income distribution. For this purpose, we explore the properties of an established model for income growth that includes ``resetting'' as a stabilising force which ensures stationary dynamics.  We find that the dynamics of inequality is regime-dependent and may range from a strictly non-ergodic state where this phenomenon has an increasing trend, up to a stable regime where inequality is steady and the system efficiently mimics ergodic behaviour. Mobility measures, conversely, are always stable over time, but the stationary value is dependent on the regime, suggesting that economies become less mobile in non-ergodic regimes. By fitting the model to empirical data for the dynamics of income share of the top earners in the United States, we provide evidence that the income dynamics in this country is consistently in a regime in which non-ergodicity characterises inequality and immobility dynamics. Our results can serve as a simple rationale for the observed real world income dynamics and as such aid in addressing non-ergodicity in various empirical settings across the globe.
\end{abstract}




\section{Introduction}
Recent studies suggest that we are currently living in a world where disparities between individual incomes are rising, and the possibilities for an individual to change their ranking in the income ladder become progressively reduced ~\cite{piketty2003income,milanovic2012global}. This phenomenon has been famously illustrated through the Great Gatsby curve, which highlights that income inequality and immobility are positively related~\cite{corak2013income}. 

A large number of studies have attempted to provide explanation for this observation. For instance, Aoki and Nirei~\cite{aoki2017zipf} investigated the impact of tax changes on the income dynamics in the United States, whereas Gabaix et al.~\cite{gabaix2016dynamics} emphasised the role of discrepancies in individual capabilities. The theoretical part of these analyses are grounded on the assumption that the individual income follows geometric Brownian motion with \textit{stochastic resetting} (srGBM). Stochastic resetting is a mechanism such that a given stochastic process evolves freely during a given random interval of time at the end of which the process is reset to its initial position \cite{evans2011diffusion,evans2020stochastic}. srGBM is a baseline model for random multiplicative income growth incorporating resetting of individual income as a stabilising force that ensures stationary dynamics. Stationarity has been a hallmark property induced by resetting as has been shown in a steadily growing number of studies recently \cite{evans2011diffusion,evans2020stochastic,gupta2014fluctuating,majumdar2015dynamical,pal2015diffusion,tal2020experimental,pal2016diffusion,ralf-ergodicity-resetting,cherstvy2021jcomplex}. The overall dynamics of srGBM, despite being stationary, remains  \textit{non-ergodic} (ensemble and long-time averages are not equivalent \cite{metzler2014anomalous}), in the sense that there might be significant differences between the observed mean (per capita) income in the population and the typical time-averaged income dynamics of individuals in the population (approximated by the median income). Recently, it was shown that the extent to which non-ergodicity affects the income dynamics is strongly dependent on the relationship between the resetting rate and the other model parameters \cite{stojkoski2021geometric}. Concretely, it was shown that the realisation of non-ergodicity in srGBM is manifested in three different regimes: i) a frozen regime in which the mean income grows infinitely, whereas the typical income converges; ii) an unstable regime, where the mean income is highly volatile and larger but convergent as the typical income; and iii) a stable regime in which there are no significant differences between the dynamics of the per capita and the typical income and the system always mimics ergodic behaviour. Non-ergodicity ultimately dictates the impact of randomness (or essentially luck) on the income dynamics of an individual. In this context, two important questions arise: i) To what degree does the randomness impact the observed inequality/mobility within an economy in the different regimes? ii) If we assume srGBM dynamics, in which regime do we currently live? 

As a means to answer these questions, we here analyse theoretically and evaluate empirically the dynamical behaviour of inequality and mobility measures, and hence quantify the Great Gatsby curve in srGBM. We find that the dynamics of three standard inequality measures that we explore: the Gini coefficient, the Share of income held by the top X\% and the Theil inequality index are also regime dependent. In frozen regime inequality displays a trend of constant growth: on the long run all the income is owned by one individual. This is a sheer result of the impact of randomness on the non-ergodic dynamics, and not an outcome of the differences between the capabilities of the individuals. In the unstable regime, inequality reaches a stationary state but its magnitude is large and highly volatile over time. As a consequence, we may observe situations in which the majority of the income is concentrated on few individuals as well as circumstances where it is fairly distributed. This is again a result of the ever-growing randomness. Finally, in the stable state, the inequality measures follow steady dynamics and assume their lowest magnitude. Conversely, the dynamics of the mobility measures that we investigate (the Spearman rank correlation and the Earnings Elasticity) are independent of the specific regime. In each case, they are stable and easily predictable observables over time. Altogether this implies that the Great Gatsby curve also features different regimes due to the dynamics of inequality. While in the stable regime, the direct relationship between inequality and mobility persists, as we move towards the unstable regime the relationship becomes highly fluctuating  because of the instability of inequality measures. In the frozen regime, inequality is at its maximum, and the system is immobile. 

We then utilise United States data for the dynamics of the income share of the top earners taken from the World Inequality Database, to study the evolution of income under the assumption that it undergoes srGBM dynamics~\cite{alvaredo2020distributional}. We find robust evidence which indicates that the United States economy is consistently in the frozen regime. Thus, we hypothesise that the economy may be in a situation where inequality is increasing and mobility is decreasing as a result of the high level of randomness that determines the income dynamics. The economics literature presented above has predominantly focused on studying the implications created by the first two regimes. There non-ergodicity does not significantly affect the income dynamics. Therefore, the studies have focused their attention to institutional and social characteristics. To this end, our results can serve as an alternative but a simple explanation for the observed real world income dynamics. Indeed, a growing body of literature suggests that, by carefully addressing the question of ergodicity, many conundrums besetting the current economic formalism can be resolved in a natural and empirically testable way~\cite{peters2019ergodicity}. 

The rest of the paper is structured as follows. In Section~\ref{sec:model} we present the srGBM model. In Section~\ref{sec:model-properties} we describe its properties and analyse the behaviour of inequality and mobility measures in it, both analytically and numerically. In Section~\ref{sec:empirical-analysis} we conduct the empirical analysis. We conclude and discuss the implications of our results in Section~\ref{sec:conclusion}.

\section{Geometric Brownian motion with stochastic resetting as a model for income dynamics 
}
\label{sec:model}

\subsection{Preliminaries}

We assume that time is continuous, and there is a continuum of workers. The income $x(t)$ of an individual worker in period $t$ follows a geometric Brownian motion with stochastic resetting (srGBM). That is, it grows multiplicatively with a rate $\mu$ and volatility $\sigma$ until a random event resets its dynamics~\cite{cherstvy2017time}. The reset event can be interpreted as a worker that left the job market (for example by retiring or being laid off) randomly with a rate $r$ and is substituted by another younger worker with a starting income $x_0 = 1$. The dynamics of the income can be described by the following Langevin equation
\begin{align}
d x(t) &=(1-Z_{t}) x(t) \left[ \mu dt+ \sigma dW \right]+Z_{t} \left( x_0-x(t) \right),
\label{eq:srgbm-microscopic}
\end{align}
where $dt$ denotes the infinitesimal time increment and
$dW$ is an infinitesimal Wiener increment, which is normal variate with $\langle dW_t \rangle=0$ and $\langle dW_t dW_s \rangle =\delta(t-s)dt$. Here $\delta(t)$ denotes the Dirac $\delta$-function. Moreover, $Z_t$ is a random variable which resets the income dynamics to the initial value $x_0$. 
$Z_t$ takes the value $1$ when there is a resetting event in the time interval between $t$ and $t+dt$; otherwise, it is zero. 

The solution to Eq.~\eqref{eq:srgbm-microscopic} can be found by interpreting the srGBM as a renewal process: each resetting event renews the income dynamics at $x_0$, and between two such consecutive renewal events the income of an individual undergoes simple GBM. Thus, between time points $0$ and $t$, only the last resetting event, occurring at the point 
\begin{align}
    t_{l}(t) = \max_{k \in \left[0,t\right]} k: \{ Z_{k} = 1 \},
    \label{eq:srgbm-solution-1-0}
\end{align}
is relevant, and the solution to Eq.~\eqref{eq:srgbm-microscopic} reads (following It\^{o} interpretation) \cite{mantegna1999introduction,bouchaud2003theory,fouque2000derivatives}
\begin{align}
      x(t) &= x_0~ e^{ (\mu - \frac{\sigma^2}{2}) \left[ t - t_{l}(t) \right] + \sigma \left[ W(t) - W(t_{l}(t)\right]}.
      \label{eq:srgbm-solution-2-0}
\end{align}
The probability for a reset event is given by $P(Z_{t} = 1) = rdt$. In the limit when $dt \to 0$, this corresponds to an exponential resetting time density $f_r(t)=re^{-rt}$, and $t_l$ is distributed according to
\begin{align}
f(t_{l}|t)=\delta(t_{l}) e^{-rt}+re^{-r(t-t_{l})},
\label{last-time-pdf}
\end{align}
such that $\int_0^t ~dt_l f(t_l|t)=1$.
Intuitively, the first term on the RHS corresponds to the scenario when there is no resetting event up to time $t$ while the second one accounts for multiple resetting events. 

\subsection{Stationary distribution}

In the literature, resetting is described as an approximation for the external forces that influence the income dynamics and ensure a stationary distribution~\cite{evans2011diffusion,evans2020stochastic,pal2015diffusion,dahlenburg2021stochastic}. One can use the renewal approach as described above (also see \cite{evans2020stochastic}) to show that that the probability density function (PDF) has a stationary solution.
In particular, the PDF with resetting $(r>0)$ can be written as 
\begin{align}\label{eq:pdf-solution resetting}
  P_r(x,t|x_0) &= e^{-rt}P_{0}(x,t|x_0)+r\int_{0}^{t}e^{-ru}P_{0}(x,u|x_0)\,du,
\end{align}
where $P_{0}(x,t|x_0)$ is the PDF of the reset-free ($r=0$) income dynamics~\cite{aitchison1957lognormal,stojkoski2020generalised}
\begin{align}
P_{0}(x,t|x_0)&= \frac{1}{x\sqrt{2\pi \sigma^2 t}}  \exp \left( \frac{-\left[\log (\frac{x}{x_0})-(\mu-\frac{\sigma^2}{2})t\right]^2}{2\sigma^2 t}  \right).
\end{align}
By Laplace transform, $\mathcal{L}[f(t)]=\int_{0}^{\infty}e^{-st}f(t)\,dt=\hat{f}(s)$, of Eq.~(\ref{eq:pdf-solution resetting}), and by using the limit $s\rightarrow0$, we find the steady state,
$P_r^{ss}(x|x_0)=\lim_{t \to \infty} P_r(x,t|x_0)=\lim_{s \to 0} s\hat{P}_r(x,t|x_0)=r \hat{P}_{0}(x,r|x_0)$. 
Following this, it can be shown that the stationary distribution follows the power law 
\begin{align}\label{solution resetting long time}
     P_r^{ss}(x|x_0) =
     \frac{r \sigma^2}{\alpha \sigma^2 + \left(\mu - \frac{\sigma^2}{2}\right)}\left\lbrace\begin{array}{l l l}
     \smallskip & \left( \frac{x}{x_0}\right)^{-\alpha-1}, \quad & x>x_0, \\ 
     & \left( \frac{x}{x_0}\right)^{\alpha+2\left(\mu - \frac{\sigma^2}{2}\right)-1}, \quad & x\leq x_0,
\end{array}\right.
     \end{align}
where
\begin{align}
    \alpha &= \frac{-(\mu - \sigma^2/2) + \sqrt{(\mu-\sigma^2/2)^2 + 2r \sigma^2}}{\sigma^2},
\end{align}
is the shape parameter.
The power law property is an important stylised fact that is prevalent in real-world income distributions~\cite{eliazar2020power,aydiner2019money}. Other stylised facts that are recovered by the model are: larger $\mu$ (larger average population growth), larger $\sigma$ (more randomness in the dynamics) and/or smaller $r$ (less retiring), result in a smaller  shape parameter and a heavier-tailed distribution. This leads to higher inequality and lower mobility in the economy. Thus, srGBM is a minimal model that is able  to adequately represent a range of real word situations. As such it has been implemented to date in various empirical studies~(see for example~\cite{gabaix2016dynamics}).

\section{Model properties}
\label{sec:model-properties}

\subsection{Moments and log-moments}

The behaviour of the inequality and mobility measures are crucially determined by the moments and the log moments of srGBM. Therefore, we begin the analysis by providing a short overview of their properties.


\paragraph{Moments:} The $n$th moment of the process can be easily derived by raising Eq.~(\ref{eq:srgbm-solution-2-0}) to the power of $n$ and then applying the law of total expectation. In general, the $n$th moment will diverge over time as long as $r < r_n \equiv n\mu + n (n-1) \frac{\sigma^2}{2}$, and it will converge to $r/\left(r-r_m\right)$ whenever $r < r_n$ \cite{stojkoski2021geometric}.
For example, by setting $n = 1$, we recover the first moment
\begin{align}
  \langle  x(t) \rangle &= \frac{\mu e^{(\mu-r)t} - r}{\mu - r}.
  \label{eq:srgbm-mean}
\end{align}
In a large enough population, the first moment is a fair approximation for the per capita income $\langle x(t) \rangle_N = \frac{\sum_i x_i(t)}{N}$, an ubiquitously used measure for economic performance~\cite{adamou2020two}. Obviously, when $r < \mu$ the income per capita will grow over time with an exponential rate $\mu-r$, whereas when $r > \mu$ it will converge to $r/(r-\mu)$.

\paragraph{Log-moments:} The $n$th log-moment can be found by implementing the same procedure, just before raising to the power of $n$, one first takes the logarithm in Eq.~(\ref{eq:srgbm-solution-2-0}). Differently from the ordinary moments, the log-moment are convergent values. For instance, the first log-moment is
\begin{align}
\langle \log x(t) \rangle &= \frac{\mu - \sigma^2/2}{r} (1 - e^{-rt})
\label{eq:srgbm-log-mean}
\end{align}
and in the long time limit it converges to $(\mu - \frac{\sigma^2}{2})/r$. We point out that the exponential of the first log-moment is the geometric mean income, and when income follows multiplicative dynamics it is a fair estimate for the typical or median income, i.e., the income which, when the individuals are ordered from the one with the lowest income to the top income earner, separates the population into two equal halves~\cite{adamou2020two}. The apparent discrepancies between the mean and the median enforce regimes in the income dynamics which ultimately affect the extent of inequality and mobility in the economy. These are discussed next.

\subsection{Regimes in srGBM}

srGBM is characterised by three physical long time \textit{non-ergodic} regimes that affect the dynamics of the per capita income in relation to the typical, median income in the population~\cite{stojkoski2021geometric}. They are: i) a frozen state regime, ii) an unstable annealed regime, and iii) a stable annealed regime. The regimes appear because the per capita income is a ``plutocratic'' measure, i.e., the individuals with larger income also play a larger role in its value and therefore in certain cases the mean income may fail to depict the behaviour of the typical individual in the population. Indeed, the regime which is observed in reality depends on the relation between the resetting rate and the other parameters of the model. 

\begin{figure}[ht!]
\includegraphics[width=14cm]{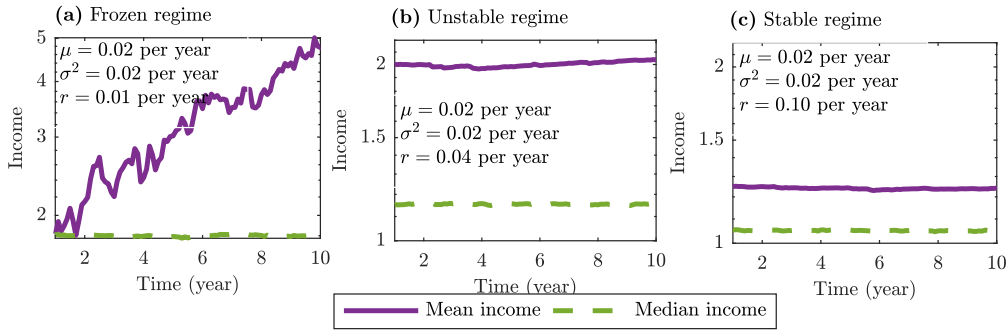}
\caption{\textbf{Regimes in srGBM.}
\textbf{(a)} Mean income and median income as a function of time in the frozen regime, \textbf{(b)} unstable regime, and \textbf{(c)} the stable regime. In \textbf{(a)-(c)} the population size is $N = 10^4$ and the initial conditions $x(0)$ are drawn from the stationary distribution. \label{fig:srgbm-regimes}}
\end{figure}

\paragraph{Frozen regime:} The first regime, named after in analogy to the celebrated ``Random Energy Model'' by Derrida~\cite{derrida1981random,fyodorov2008freezing}, appears when $\mu > r$ (alternatively, when $\alpha < 1$). In this case, all the moments are divergent. Thus, for a large population, the mean income $\langle x \rangle_N$  will increase indefinitely, whereas the median income will saturate at a finite value. This is because resetting occurs at a lower rate than the growth of income. As a consequence, the per capita income on the long run is dominated by a small number of individual workers whose income is yet to reset and will be significantly larger than the median income (see Fig.~\ref{fig:srgbm-regimes}(a)).

\paragraph{Unstable regime:} The second regime is observed when $2\mu+\sigma^2 > r > \mu$ (or when $1<\alpha < 2$). Then, the first moment of srGBM is convergent, but the variance is divergent. Therefore, in this situation we will observe a mean income that is apparently stationary. However, due to the divergent variance, the mean income will be an unstable quantity over time and it will be notably different when compared to the median income (Fig.~\ref{fig:srgbm-regimes}(b)).

\paragraph{Stable regime:} Finally, the last regime appears when $2\mu+\sigma^2 < r$ (or $\alpha>2$). In this case, the variance of srGBM is convergent and there are no significant differences between the stationary dynamics of the median and mean income (Fig.~\ref{fig:srgbm-regimes}(c)).

\subsection{Inequality in srGBM}

Let us now turn our attention to one of the main topics of this study and examine how income inequality measures behave in the different srGBM regimes. Formally, income inequality can be seen as a static concept. It quantifies the extent of concentration in the income distribution. In this context, an inequality measure is a function that ascribes a value to the observed distribution of income in a way that allows direct and objective comparisons across different economies and points in time~\cite{afonso2015inequality,stojkoski2019cooperation,stojkoski2021evolution}. In situations when the income is concentrated on a small number of individuals the measure will suggest larger inequality, and vice versa, when the income is more spread across the population it will suggest a more egalitarian society. In the economics literature, a variety of income inequality measures have been developed (for a review, see Ref.~\cite{eliazar2018tour}). In what follows we will analyse the underlying features of three most widely used empirical measures by assuming srGBM dynamics: the Gini coefficient, the share of income owned by the top X\% individuals in the population and the Theil inequality index.

\paragraph{Gini coefficient:} The Gini coefficient quantifies the extent to which the observed income distribution differs from the line of perfect equality, i.e., the income distribution in a hypothetical society where every individual has the same income \cite{eliazar2010maximization}. Hence, this measure is uniquely identified by the shape of the PDF of the stationary income dynamics. Mathematically, given the cumulative distribution function $F(x,t|x_0) = \int_0^x P_r(x',t|x_0)dx'$ of the stationary income, the Gini coefficient can be expressed as
\begin{align}
    G(t) &= \frac{1}{\langle x(t) \rangle} \int_0^\infty F(x,t|x_0) \left(1 - F(x,t|x_0) \right) dx.
\end{align}
From the above equation, it can be seen that the Gini coefficient is a normalised measure whose values are between 0 and 1, with larger values indicating larger income inequality \cite{eliazar2010maximization}.

Traditionally, changes in the income distribution are expressed through changes in model parameters, reflecting shocks in economic conditions, with rapid equilibration thereafter. Therefore, studies of income inequality are usually interested in the behaviour of the Gini coefficient which occurs in the stationary state of srGBM. We already know that the model is characterised in a power law distribution. The theoretical behaviour of the Gini coefficient in this type of distribution is approximately $1$ if $\alpha < 1$ and
$1/(2\alpha-1)$ otherwise~\cite{dorfman1979formula,eliazar2014social}. In other words, in the frozen regime the majority of the income will be owned by one individual, whereas out of the frozen regime inequality will decrease proportionally to the shape parameter of the stationary distribution. Empirically, however, the behaviour will be slightly different. This is because even though theoretically the moments of the process are allowed to be infinite, the system consists of a finite population, and the moments of the process will always be a fixed finite value that can be divergent with a speed dependent on the economic conditions. 

\begin{figure}[ht!]
\includegraphics[width=14cm]{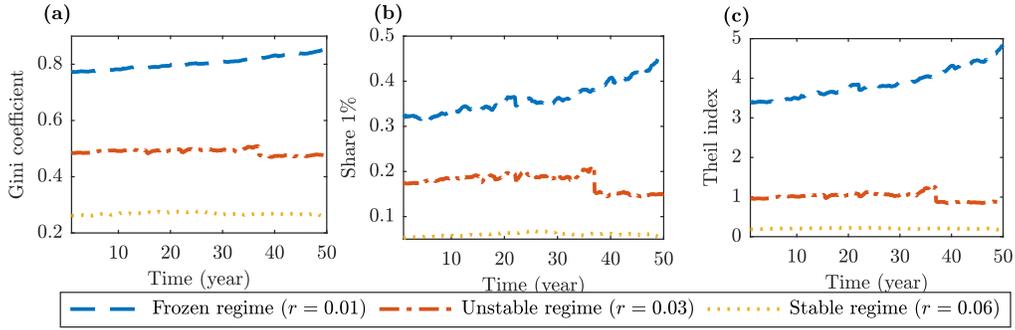}
\caption{\textbf{Inequality measures in srGBM.}
\textbf{(a)} Gini coefficient versus time in the stationary state of srGBM. \textbf{(b)} Same as \textbf{(a)}, only for the Share of income held by the top 1\%. \textbf{(c)} Same as \textbf{(a)} only for the Theil index. \textbf{a-c} We set $N = 10^4$, $\mu = 0.02$ per year and $\sigma^2= 0.02$ per year. The initial conditions $x(0)$ are drawn from the stationary distribution. \label{fig:srgbm-inequality}}
\end{figure}

To illustrate the empirical behaviour of the Gini coefficient, we numerically simulate the income dynamics in an economy consisting of $N = 10^4$ individuals. The results are depicted in Fig.~\ref{fig:srgbm-inequality}(a) where we plot the Gini coefficient as a function of time in the three different regimes of srGBM. In the frozen regime of srGBM, inequality as measured through the Gini coefficient exhibits a growing trend, and it is converging towards its theoretical value (maximum inequality). In the unstable regime, the Gini coefficient reaches a stationary value that is close to its theoretical estimate. Nonetheless, because the variance of the income is divergent, randomness plays a large role in the dynamics. Finally, in the stable regime we observe dynamics that are essentially stable, as it should be.

\paragraph{Share of income held by the top X\%:} The power law property describes the fractal nature of the income distribution. That is, the top 0.1\% of the individuals with the highest income are X times richer on average than the top 1\% who are, in turn, X times richer than the top 10\%, where X is a fixed number. As a result, the share of income held by the top X\% has been usually implemented as a measure that quantifies the concentration of the income in the tail of the distribution. As in the case with the Gini coefficient, this measure also lies between $0$ and $1$ with larger values suggesting higher inequality. Moreover, as can be seen in Fig.~\ref{fig:srgbm-inequality}(b), its dynamical behaviour is exactly the same as the Gini coefficient. Concretely, we show the evolution of the Share of the top 1\% as a function of time for the same finite population as in Fig.~\ref{fig:srgbm-inequality}(a). We observe that the Share increases indefinitely in the frozen regime, followed with highly volatile dynamics in the unstable state, and steady and stationary dynamics in the stable state.

\paragraph{Theil inequality index:} We now examine the Theil inequality index which is estimated as
\begin{align}
    \mathrm{Th}(t) &= \log \langle x(t) \rangle - \langle \log x(t) \rangle.
    \label{eq:theil}
\end{align}
Basically, this index is the difference between the logarithms of the mean income and the mean of the logarithmic incomes of the population, i.e., the difference of Eq.~\eqref{eq:srgbm-mean} and Eq.~\eqref{eq:srgbm-log-mean}. As a result, unlike the previous two measures, the Theil index is unbounded from above and as the inequality in the economy increases, it diverges towards infinity. Recent studies suggest that this quantity naturally arises as a unique inequality measure in economic systems in which the income undergoes multiplicative dynamics~\cite{adamou2020two}.

Obviously, the empirical behaviour of the index when income follows srGBM dynamics is determined by the per capita income: the Theil index will diverge towards infinity in the frozen regime, will be convergent but at a highly uncertain value in the unstable regime and it will exhibit steady dynamics in the stable regime. This is again similar to the behaviour of the Share of income held by the top individuals and the Gini coefficient, and is summarised in Fig.~\ref{fig:srgbm-inequality}.

In summary, each of the studied inequality measures above exhibits identical qualitative behaviour, thus suggesting that the regime dependence is a robust feature in the income dynamics.

\subsection{Mobility in srGBM}

Measures of economic mobility quantify how income ranks of individuals change over time. Intuitively, when mobility is high, the chances of an individual to change their position in the income distribution over a given time period are high. 
In contrast, when mobility is low, individuals are unlikely to change their rank in the distribution over time, or the changes may happen slowly. As a means to study the drivers of economic mobility in srGBM we consider two standard measures: Spearman's rank correlation and the earnings elasticity (EE). In fact, both the rank correlation and the EE are measures of immobility, and to consider them as measures of mobility one has to consider their complement or their inverse. 

\paragraph{Spearman's rank correlation:} Spearman's rank correlation $\rho_{t,\Delta}$is defined on a joint distribution of income at two points in time, $t$ and $t+\Delta$. Mathematically, it reads
\begin{align}
    \rho_{t,\Delta} = 1 - \frac{6\sum_i \left[rg\left(x_i\left(t\right)\right) - rg\left(x_i\left(t + \Delta \right)\right)\right]^2}{N\left(N^2-1\right)}\,,
\end{align}
where $rg(x)$ is the rank transformation of $x$. This measure is bounded between $-1$ and $1$. $\rho_{t,\Delta} = 1$ suggests perfect immobility, a state in which there is no change in income ranks between the two points in time. Lower values suggest greater economic mobility. 

\begin{figure}[ht!]
\includegraphics[width=14cm]{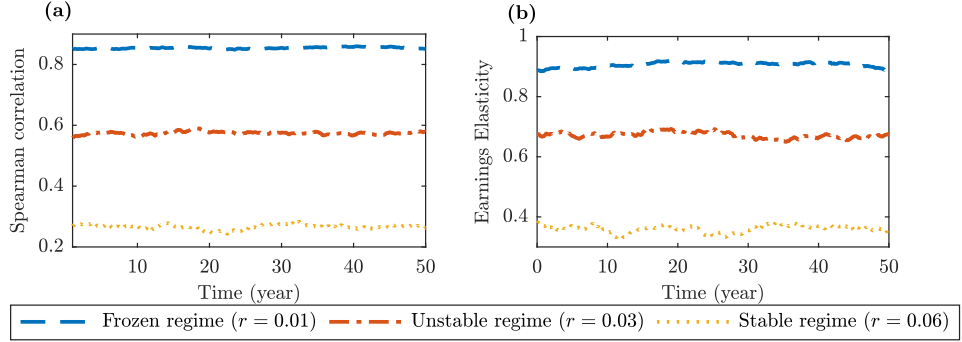}
\caption{\textbf{Mobility measures in srGBM.}
\textbf{(a)} Spearman rank correlation versus time in the stationary state of srGBM. \textbf{(b)} Same as \textbf{(a)} only for the Earnings elasticity. 
\textbf{a-b} The population size is $N = 10^4$, $\mu = 0.02$ per year, $\sigma^2= 0.02$ per year and the initial conditions $x(0)$ are drawn from the stationary distribution. The temporal difference $\Delta$ is 10 years.  \label{fig:srgbm-mobility}}
\end{figure}

Interestingly though, differently from the inequality measures, the srGBM regime does not impact the dynamics of the Spearman rank correlation. As can be seen in Fig.~\ref{fig:srgbm-mobility}(a), where we numerically simulate an economy in the three different regimes, the regimes only have an impact on the mean value of the rank correlation: its dynamics are stable around this value.

\vspace{0.5cm}
\paragraph{Earnings elasticity:} The Earnings Elasticity is defined as the slope $b_{t,\Delta}$ of the regression
\begin{align}
   \log\left(x_i\left(t+\Delta\right)\right) = b_0 + b_{t,\Delta} \log\left(x_i\left(t\right)\right) + u_i\,,
\end{align}
where $b_0$ is the intercept and $u_i$ is the error term. This is a simple linear regression and therefore,
\begin{align}
    b_{t,\Delta} = \mathrm{corr}\left[\log\left(x\left(t\right)\right),\log\left(x\left(t+\Delta\right)\right)\right] \frac{\mathrm{var}\left[\log\left(x\left(t+\Delta\right)\right)\right]}{\mathrm{var}\left[\log\left(x\left(t\right)\right)\right]}\,,
    \label{eq:iee-estimation}
\end{align}
where $\mathrm{corr}(x,y)$ is the correlation, between the variables $x$ and $y$, i.e., 
\begin{align}
    \mathrm{corr}\left[x, y\right] &= \frac{\mathrm{cov}[x, y]}{\sqrt{\mathrm{var}[x]} \sqrt{\mathrm{var}[y]}},
    \label{eq:autocorrelation}
\end{align}
with 
\begin{align}
\mathrm{cov}[x, y] &\equiv \langle x y \rangle -\langle  x \rangle \langle y  \rangle 
\label{eq:bm_restart_covariance}
\end{align}
being the covariance of the same variables and $\mathrm{var}(x)$ is the variance of $x$. As with the rank correlation, lower EE also indicates greater mobility. However, this measure is unbounded and may take on any real values. 

In the stationary regime of srGBM the EE can be analytically quantified by knowing that the logarithm of income follows a standard diffusion with stochastic resetting that has a drift~\cite{majumdar2018spectral,stojkoski2021autocorrelation}. In particular, because of the stationarity, it follows that $\mathrm{var}\left[\log\left(x\left(t\right)\right)\right] = \mathrm{var}\left[\log\left(x\left(t+\Delta\right)\right)\right]$ and therefore $b_{t,\Delta}$ is simply given by the autocorrelation function of diffusion with stochastic resetting. Mathematically, we have
\begin{align}
   b_{t,\Delta} &=  \mathrm{corr}\left[\log\left(x\left(t\right)\right),\log\left(x\left(t+\Delta\right)\right)\right] = e^{-r\Delta}.
\end{align}
For the derivation of the analytical expression of the autocorrelation function we refer to~\cite{stojkoski2021autocorrelation}. Similarly to the properties of the rank correlation, and as depicted in~Fig.~\ref{fig:srgbm-mobility}(b), the dynamics of EE are also independent on the regime.

\subsection{The Great Gatsby curve in srGBM}

Quantifying the behaviour of the various inequality and mobility measures in srGBM allows us to study the relationship between them in an economy. A standard approach for visualising the relationship is through the Great Gatsby curve. The curve was introduced by Corak~\cite{corak2013income} who utilised country level data to illustrate the association between the Gini coefficient and the Earnings Elasticity, indicating that economic immobility and inequality are \textit{positively related}, implying that economies with a more unequal distribution of income also offer fewer perspectives for individuals to change their income rank. 

\begin{figure}[ht!]
\includegraphics[width=12cm]{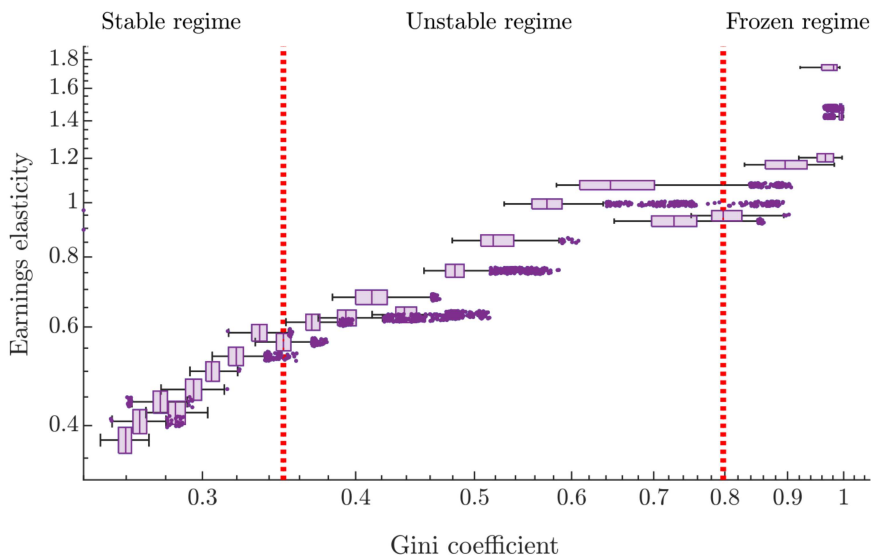}
\caption{\textbf{The Great Gatsby Curve in srGBM.}
Boxplots for the stationary Gini coefficient for various various choices of the stationary Earnings elasticity. The results are averaged across $10^3$ realisations and the median results are shown. The violet filled scatters are the extreme outliers. The population size is $N = 10^4$, $\mu = 0.02$ per year, $\sigma^2= 0.02$ per year and the initial conditions $x(0)$ are drawn from the stationary distribution. The resetting rate ranges from $0$ up to $0.12$ per year and the temporal difference $\Delta$ is 10 years. \label{fig:srgbm-ggatsby}}
\end{figure}

In Fig.~\ref{fig:srgbm-ggatsby} we display numerical results for the Great Gatsby curve by simulating the srGBM in the stationary state $10^3$ times and afterwards quantifying the boxplots for the distribution of the generated Gini coefficients and Earnings Elasticity for various resetting rates. The figure demonstrates that the model adequately reproduces the qualitative empirical observation that inequality and immobility are positively related. However, the regime dependence of the inequality measures is mirrored in the qualitative behaviour of the Great Gatsby curve. While in the stable regime, it is easy to notice the direct relationship between the Gini coefficient and the Earnings Elasticity, the direction of the relationship becomes unclear in the unstable regime. This is because then inequality is a highly volatile phenomenon and is uniquely determined by the degree of randomness in the system. In the frozen regime, the Gini coefficient reaches its maximum value, and then the direct relationship between inequality and immobility vanishes.

\section{Empirical analysis}
\label{sec:empirical-analysis}

The results presented in the previous section lead to important implications regarding the dynamics of inequality and mobility within an economy. This, in turn, should impact the implementation of socio-economic policies that are aimed at improving the overall welfare of the society. More precisely, the stable regime can be seen as a Utopian situation in which inequality and immobility are easily related and one can use standard welfare policies to reduce their impact on the society. However, if an economy is in the unstable regime, then the observed inequality should not be taken for granted as it is a highly volatile variable. In this case it is the stability of mobility that allows for each individual to eventually experience the different positions in the income rankings. Moreover, if the system is in the frozen regime, then it is essentially meaningless to follow ideologies under which individuals are valued by their ability or achievement, as then it is essentially luck that determines the position of an individual in the income ranks. 

A simple approach that can be used for evaluating empirically the regime in which an economy currently is, is to quantify the shape of its income distribution. Indeed, there have been plenty of such studies. The general consensus from these studies is that income distributions are usually best described with shape parameters that suggest that economies are in the unstable regime (see, for example, Ref.~\cite{druagulescu2001exponential} for a study about the income distributions in United States and United Kingdom and~\cite{clementi2005power} for Italy). 

The weakness of this approach is that it does not take into account the potential dynamics that result from changes in the system conditions. An alternate approach, which accounts for dynamics has been recently introduced in~\cite{berman2020wealth} to tackle the dynamical behaviour of the wealth distribution in the United States. In what follows we will utilise the same procedure to model the income dynamics in the same country.

\subsection{Method}

We assume that the income dynamics follows srGBM that is constantly under the threat of changing its parameters. In this context, we will assume that the resetting rate $r(t)$ to be estimated is a function of time and will provide an approximation $\hat{r}(T)$ with the fraction of people that lost and/or left their job. For simplicity, we will measure the resetting rate on a yearly basis and assume that in between two years the resetting rate is fixed, i.e., $r(t) \approx \hat{r}(T)$ for any $t$ between $T$ and $T+1$. Our goal is to simultaneously provide consistent estimates $\hat{\mu}(T)$ and $\hat{\sigma}(T)$ for the drift parameter $\mu(T)$ and the noise amplitude $\sigma(T)$ as a function of the time, that best fits the observed Shares of income owned by the top 1\% and the top 10\% in the US income distribution. The assumption for dynamics in the model parameters reflects the possibility of noise in the data. In addition, it can be an approximation for the changes in economic conditions that affect the srGBM dynamics. These can be either due to changes in government policies or due to circumstances that are not under the control of the policymakers.



Formally, the estimation procedure consists of the following steps: 

\textbf{Step 1}: Fix the resetting rate $\hat{r}(0)$ in the initial period at the initial year $T=0$ and then estimate $\hat{\mu}(0)$ and $\hat{\sigma}(0)$ to match the srGBM stationary distribution.

\textbf{Step 2}: Propagate $N$ individual income trajectories according to the laws of srGBM. That is, with probability $1-\hat{r}(T)\Delta t$ the income undergoes GBM so that: 
\begin{align}\label{simulations}
x_i(t+\Delta t) = x_i(t) + x_i(t)[\hat{\mu}(T) \Delta t + \hat{\sigma}(T)\sqrt{\Delta t}\eta_i(\Delta t)],
\end{align} 
where $\eta_i(\Delta t)$ is a Gaussian random variable with zero mean and unit variance, and $\Delta t$ is a small time increment. With complementary probability $\hat{r}(T)\Delta t$, the income resets to the initial position:
\begin{align}\label{simulations2}
x_i(t+\Delta t) = 1.
\end{align}
At last, we find the values $\hat{\mu}(T+1)$ and $\hat{\sigma}(T+1)$ that minimise the squared difference between the inferred share of the top 1\% (or top 10\%) in the modelled population in year $T+1$ and the observed share in real data.

\textbf{Step 3}: Repeat Step 2 until the end of the time series.

For each time series we run a simulation for a model economy of $N=10^6$ workers. However, because of the randomness of the numeric simulations, each simulation will result in different fitted values. To take this into consideration, we construct a Monte Carlo estimation by repeating the process 100 times and report the average value of $\hat{\mu}(T)$ and $\hat{\sigma}(T)$. In addition, this allows us to estimate the variability of the results and provide confidence intervals for both parameters.

\subsection{Data and interpretation of the resetting rate}


Each year, the true resetting rate $r(t)$, is approximated with the fraction of the working age population (15-64) in USA who lost and/or left their job within a calendar year. The people who lost their job are those that either are temporarily laid off as well as those who permanently lost their jobs. The job leavers, on the other hand, are those that quit and immediately began searching for a new work. We take these data from the dataset for unemployment provided by the U.S. Bureau of Labor Statistics. The time series covers the period from 1977 up to 2015 and can be accessed at~\url{ https://fred.stlouisfed.org}.


The data for the top 1\% and top 10\% used for assessing the robustness were taken from the The World Inequality Database (\url{https://wid.world/}).

\subsection{Results}

\begin{figure}[ht!]
\includegraphics[width=12.5cm]{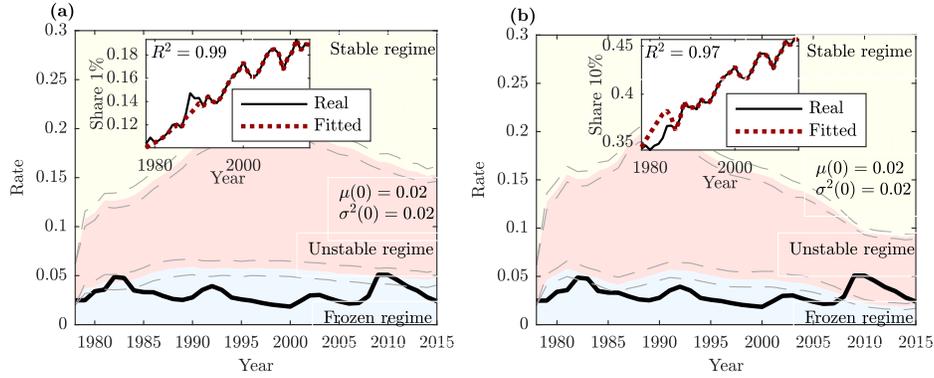}
\caption{\textbf{srGBM dynamics in the United States.}  Resetting rate $r$ as a function of time (black line) that best fits the Share of top 1\% (\textbf{(a)}) and the Share of top 10\% (\textbf{(b)}). \textbf{(a-b)} The region is coloured according to the estimated thresholds for the regimes. The dashed lines are estimated 95\% confidence intervals for the regime thresholds. The inset plots provide the real and fitted shares. The data for the income share are taken from the World Inequality Database and the resetting rate data are from the U.S. Bureau of Labor Statistics. \label{fig:srgbm-united states}}
\end{figure}

The empirical results are shown in Fig.~\ref{fig:srgbm-united states}. The figure displays the inferred regions of the regimes based on the fitted values $\hat{\mu}(T)$ and $\hat{\sigma}(T)$, together with the approximation of the resetting rate as a function of time for the share of the top 1\% (Fig.~\ref{fig:srgbm-united states}(a)) and the share of the top 10\% (Fig.~\ref{fig:srgbm-united states}(b)). The inset plots provide the fitted and the shares observed in reality. It can be seen that in both cases, the fitted share accounts for more than $95\%$ of the observed cross period variation, (the value of the \textit{coefficient of determination} $R^2$
in each case is above $0.95$), and therefore it can be argued that the model adequately predicts the income dynamics in the United States. More importantly, when we look at the evolution of the regimes in the country, we find that the fitted growth rate of income is persistently above the resetting rate $\hat{r}(T)$ (black line) in both inference procedures. This leads us to the conclusion that the process is robustly in the frozen regime (the highlighted blue region) for the time period considered. Only in the period after the Financial Crisis of 2010, when modelling the top 10\% we observe a change towards the unstable regime (orange region), although this change is relatively small. Hence, the evidence we present here suggests that the
US income dynamics is in a state where it is uniquely determined by the frozen regime and thus the dynamics is characterised by non-ergodicity. 

\section{Conclusion}
\label{sec:conclusion}


Various theories have been developed for explaining the observed rise in income inequality and the decline in mobility all over the world. In our study, even though it is based on these theories, we took a slightly more agnostic approach and examined the impact of non-ergodicity, i.e., randomness on the income dynamics. 

For this purpose, we utilised the properties of srGBM, an established model for non-ergodic income growth. The fact that in srGBM, the extent of non-ergodicity is manifested via different regimes, allowed us to perform a theoretical analysis for the behaviour of income inequality and mobility measures. This, in turn, aided us in devising an empirical test for investigating the presence of different regimes in real world income dynamics.

We showed that different regimes also appear in the behaviour of inequality measures, but not in the dynamics of mobility measures. That is, the dynamics of income inequality may range from a strictly non-ergodic state in which this phenomenon empirically is always increasing, up to a stable regime where inequality is steady and the system efficiently mimics ergodic behaviour. Mobility measures, on the other hand, are always stable over time, just reach a different stationary value, suggesting that economies become less mobile in non-ergodic regimes. This is eventually translated in the dynamical behaviour of the Great Gatsby curve, i.e., the visual method for describing the relationship between inequality and mobility.

The theoretical analysis was coupled with an empirical investigation for the income dynamics in the United States. The investigation unveiled that this country is persistently in a frozen non-ergodic regime where income inequality converges towards its maximal value and the possibilities for an individual to change their position on the income ladder will become even lower. Interestingly, coinciding conclusions for the dynamics of wealth were found in~\cite{berman2020wealth}. Hence, these findings allow us to conjecture that non-ergodicity may serve as a simple explanation for the observed diverging inequality and immobility in the United States.
The srGBM empirical methodology presented here is simple and offers great explanatory power. Therefore,  it would be interesting to apply this methodology to other economies and investigate the potential impact of non-ergodicity on a global level. We believe that, with the availability of novel data, this will be easily attainable in the near future.

\section*{Acknowledgments}
V.S., T.S., L.K. and R.M. acknowledge financial support by the German Science Foundation (DFG, Grant number ME 1535/12-1). T.S. was supported by the Alexander von Humboldt Foundation. R.M. acknowledges support from the Foundation for Polish Science (Fundacja na rzecz Nauki Polskiej, FNR) within an Alexander von Humboldt Honorary Polish Research Scholarship.








\end{document}